\documentclass[english,prd,onecolumn,superscriptaddress,nofootinbib,preprintnumbers]{revtex4}
\usepackage[latin1]{inputenc}
\usepackage{graphicx}
\usepackage{amsmath,amsthm,amssymb}
\usepackage{bm}

\makeatletter

\usepackage{amsfonts}
\usepackage{dcolumn}
\usepackage{hyperref}

\def\be{\begin{equation}}
\def\ee{\end{equation}}
\def\ba{\begin{eqnarray}}
\def\ea{\end{eqnarray}}
\def\bs{\begin{subequations}}
\def\es{\end{subequations}}

\def\tA{\tilde{A}}
\def\tB{\tilde{B}}
\def\tC{\tilde{C}}

\usepackage{color}

\makeatother

\usepackage{babel}
\makeatother
\begin{document}

\title{Matter perturbations in Galileon cosmology}

\author{Antonio De Felice}
\affiliation{Department of Physics, Faculty of Science, Tokyo University of Science,
1-3, Kagurazaka, Shinjuku-ku, Tokyo 162-8601, Japan}

\author{Ryotaro Kase}
\affiliation{Department of Physics, Faculty of Science, Tokyo University of Science,
1-3, Kagurazaka, Shinjuku-ku, Tokyo 162-8601, Japan}

\author{Shinji Tsujikawa}
\affiliation{Department of Physics, Faculty of Science, Tokyo University of Science,
1-3, Kagurazaka, Shinjuku-ku, Tokyo 162-8601, Japan}

\begin{abstract}

We study the evolution of matter density perturbations in
Galileon cosmology where the late-time cosmic acceleration 
can be realized by a field kinetic energy.
We obtain full perturbation equations at linear order
in the presence of five covariant Lagrangians ${\cal L}_i$ ($i=1, \cdots, 5$)
satisfying the Galileon symmetry $\partial_{\mu}\phi \to \partial_{\mu}\phi+b_{\mu}$ in the flat space-time. 
The equations for a matter perturbation as well as an effective 
gravitational potential are derived under a quasistatic approximation 
on subhorizon scales. 
This approximation can reproduce full numerical solutions
with high accuracy for the wavelengths relevant to large-scale structures.
For the model parameters constrained by the background expansion 
history of the Universe the growth rate of matter perturbations 
is larger than that in the $\Lambda$-cold dark matter model, 
with the growth index $\gamma$ today typically smaller than 0.4. 
We also find that, even on very large scales associated with 
the integrated-Sachs-Wolfe effect in cosmic microwave 
background temperature anisotropies, 
the effective gravitational potential exhibits a
temporal growth during the transition from the matter
era to the epoch of cosmic acceleration.
These properties are useful to distinguish the 
Galileon model from the $\Lambda$-cold dark matter in future
high-precision observations.

\end{abstract}

\date{\today}


\maketitle

\section{introduction}

The large-distance modification of gravity has received much attention 
as a possible explanation for the cosmic acceleration 
today \cite{review}.
Many modified gravitational models of dark energy have been already 
proposed--including those based on $f(R)$ gravity \cite{fRearly}, 
scalar-tensor theories \cite{stensor}, Gauss-Bonnet gravity and its generalizations \cite{GBearly}, 
Dvali-Gabadazde-Porrati (DGP) braneworld \cite{DGP}, and 
Galileon gravity \cite{Nicolis}. In $f(R)$ gravity, for example, the viable 
models \cite{fRviable} are designed 
to have a large mass of a scalar gravitational degree 
of freedom (``scalaron'' \cite{Star80}) in the regions of 
high density for the compatibility with local gravity experiments.
In this case, as long as the so-called chameleon 
mechanism \cite{chameleon} is at work, 
the interaction between the scalaron and baryons can be suppressed
to satisfy local gravity constraints \cite{fRlocal}.

There is another mechanism to decouple the fifth force from baryons at short distances even in the absence of the scalar-field potential.
Nonlinear effects of field self-interactions can allow the recovery 
of General Relativity (GR) inside a so-called Vainshtein 
radius \cite{Vainshtein}. In the DGP model a longitudinal graviton (i.e. a brane-bending mode $\phi$) gives rise to the self-interaction of the form 
$(r_c^2/m_{\rm pl})\,\square \phi (\partial^{\mu} \phi 
\partial_{\mu} \phi)$ through the mixing with a transverse graviton, 
where $r_c$ is a crossover scale of the order of the Hubble radius 
$H_0^{-1}$ today and $m_{\rm pl}$ is the Planck mass \cite{DGPnon}.
In the local region where the energy density $\rho$ is much larger
than $r_c^{-2}m_{\rm pl}^2$ the nonlinear self-interaction 
can lead to the decoupling of the field from matter.
However the DGP model suffers from a ghost problem \cite{DGPghost}, 
in addition to the difficulty for consistency with the combined 
data analysis of Supernovae Ia (SN Ia) and Baryon Acoustic 
Oscillations (BAO) \cite{DGPobser}.

The self-interacting Lagrangian 
$\square \phi (\partial^{\mu} \phi \partial_{\mu} \phi)$ 
appearing in the DGP model
satisfies the Galileon symmetry $\partial_{\mu} \phi \to 
\partial_{\mu} \phi+b_{\mu}$ in the Minkowski background.
Imposing the Galileon symmetry in the flat space-time
one can show that the field Lagrangian consists of five terms 
${\cal L}_1, \cdots, {\cal L}_5$,
where the term $\square \phi (\partial^{\mu} \phi \partial_{\mu} \phi)$ 
corresponds to ${\cal L}_3$ \cite{Nicolis}.
In Refs.~\cite{Deffayet} these terms were extended to 
covariant forms in the curved space-time.
Moreover one can keep the equations of motion 
up to the second-order, while recovering the Galileon 
Lagrangian in the Minkowski space-time.
This property is welcome to avoid the appearance of 
an extra degree of freedom associated with ghosts. 

In Refs.~\cite{DT2,DT3} two of the present authors studied 
the cosmological dynamics of covariant Galileon theory 
in the presence of the terms up to ${\cal L}_5$
(see Refs.~\cite{JustinGal}-\cite{Kimura} for related works).
There exist de Sitter (dS) solutions responsible for dark energy 
driven by the field kinetic energy.
Refs.~\cite{DT2,DT3} also clarified the viable model parameter 
space in which the appearance of ghosts and instabilities of 
scalar and tensor perturbations can be avoided.

In the covariant Galileon cosmology the solutions finally converge 
to a common trajectory (tracker), which is 
characterized by the evolution $\dot{\phi} \propto H^{-1}$ 
($H$ is the Hubble parameter) \cite{DT2}.
The epoch at which the solutions approach the tracker 
depends on the initial conditions of the variable
$r_1 \equiv \dot{\phi}_{\rm dS}H_{\rm dS}/
({\dot{\phi}H})$, 
where the subscript ``dS'' represents the values at the dS point.
For smaller initial values of $r_1$ the approach to 
the tracker, characterized by $r_1=1$, occurs later.
In Ref.~\cite{NFT} it was shown that the combined
data analysis of SN Ia, BAO, and the cosmic microwave background (CMB) shift parameter
tends to favor a late-time tracking behavior around
the present epoch. This comes from the fact that 
the dark energy equation of state $w_{\rm DE} \simeq -2$, 
which corresponds to the one for the tracker during the deep 
matter era \cite{DT2}, is difficult to be compatible 
with a number of observations.

In order to constrain the Galileon model further, it is important to 
know the evolution of cosmological perturbations from the matter era 
to the epoch of cosmic acceleration.  
In particular, the modified growth of matter perturbations 
$\delta_m$ affects the matter power spectrum as well as 
the weak lensing spectrum \cite{obsermo}.
Moreover the modification of gravity manifests itself for 
the evolution of the effective gravitational potential 
$\Phi_{\rm eff}$ related with the integrated-Sachs-Wolfe (ISW) effect 
in CMB anisotropies.
In this paper we shall derive the equations of cosmological perturbations 
in the presence of the five Galileon terms ${\cal L}_i$ ($i=1, \cdots, 5$)
and numerically integrate them to find observational signatures
of the Galileon model. 
We also obtain convenient forms of the effective gravitational 
coupling $G_{\rm eff}$ and the anisotropic parameter $\eta$
between two gravitational potentials, under a quasistatic 
approximation for the modes deep inside the Hubble radius.

This paper is organized as follows.
In Sec.~\ref{secback} we review the Galileon cosmology 
in the flat Friedmann-Lema\^{i}tre-Robertson-Walker (FLRW) 
background.
In Sec.~\ref{seccosmoper} the equations of
cosmological perturbations in the Galileon model
are derived in the presence of nonrelativistic matter.
In Sec.~\ref{secsubho} we obtain the equations for
$\delta_m$ and $\Phi_{\rm eff}$
under a quasistatic approximation on subhorizon scales.
In Sec.~\ref{secnume} we present numerical results for the 
evolution of perturbations by integrating the full 
equations of motion for the wave numbers relevant to 
large-scale structures and CMB anisotropies. 
Sec.~\ref{secconclude} is devoted to conclusions.

\section{Background Galileon cosmology}
\label{secback}

Let us consider the following action
\be
S=\int {\rm d}^{4}x\sqrt{-g}\left[ \frac{M_{\rm pl}^{2}}{2}R
+\frac{1}{2}\sum_{i=1}^{5}c_{i}\mathcal{L}_{i}
+p_m (\mu, s) \right] \,,
\label{action}
\ee
where $g$ is a determinant of the space-time metric $g_{\mu \nu}$, 
$M_{\rm pl}=(8\pi G)^{-1/2}$ is the reduced Planck 
mass (with $G$ being gravitational constant), 
$R$ is a Ricci scalar, and $c_i$ are constants.
The five covariant Lagrangians $\mathcal{L}_{i}$ ($i=1, \cdots, 5$) 
are given by \cite{Deffayet}
\ba
& & {\cal L}_1=M^3 \phi\,,\quad
{\cal L}_2=(\nabla \phi)^2\,,\quad
{\cal L}_3=(\square \phi) (\nabla \phi)^2/M^3\,, \nonumber \\
& & {\cal L}_4=(\nabla \phi)^2 \left[2 (\square \phi)^2
-2 \phi_{;\mu \nu} \phi^{;\mu \nu}-R(\nabla \phi)^2/2 \right]/M^6,
\nonumber \\
& & {\cal L}_5=(\nabla \phi)^2 [ (\square \phi)^3
-3(\square \phi)\,\phi_{; \mu \nu} \phi^{;\mu \nu}
+2{\phi_{;\mu}}^{\nu} {\phi_{;\nu}}^{\rho}
{\phi_{;\rho}}^{\mu}
-6 \phi_{;\mu} \phi^{;\mu \nu}\phi^{;\rho}G_{\nu \rho} ]
/M^9\,,
\label{lag}
\ea
where $M$ is a constant having a dimension of mass, and $G_{\nu \rho}$
is the Einstein tensor.  These Lagrangians respect the Galileon
symmetry in the Minkowski space-time. Moreover the field equations are
kept up to the second-order in time derivatives.
The last term, $p_m$, represents the pressure of a perfect fluid whose energy density and equation of state are $\rho_m$ and $w=p_m/\rho_m$, respectively. The pressure depends on a chemical potential
$\mu=(\rho_m+p_m)/n$ ($n$ is a number density) 
and an entropy per particle $s$.

Since we are interested in the evolution of matter density
perturbations long after the radiation-domination, we take into
account a nonrelativistic fluid ($w \simeq 0$) only in the
following discussion.  In the FLRW space-time with the scale factor $a(t)$,
the variation of the action (\ref{lag}) leads to the following background
equations \cite{DT2,DT3}:
\ba
& & 3M_{\rm pl}^2 H^2=\rho_{\rm DE}+\rho_m\,,
\label{be1} \\
& & 3M_{\rm pl}^2 H^2+2M_{\rm pl}^2 \dot{H}=-p_{\rm DE}\,,
\label{be2} 
\ea
where $H=\dot{a}/a$ is the Hubble parameter (a dot represents
a derivative with respect to cosmic time $t$), and
\ba
\hspace{-0.5cm}\rho_{\rm DE} &\equiv& -c_1 M^3 \phi/2-c_2 \dot{\phi}^2/2
+3c_3 H \dot{\phi}^3/M^3 -45 c_4 H^2 \dot{\phi}^4/(2M^6)
+21c_5 H^3 \dot{\phi}^5/M^9,\\
\hspace{-0.5cm}p_{\rm DE} &\equiv&  c_1 M^3 \phi/2-c_2 \dot{\phi}^2/2
-c_3 \dot{\phi}^2 \ddot{\phi}/M^3 
+3c_4 \dot{\phi}^3 [8H\ddot{\phi} +(3H^2+2\dot{H})
\dot{\phi}]/(2 M^6) 
-3c_5 H \dot{\phi}^4 [5H \ddot{\phi}+2(H^2+\dot{H}) 
\dot{\phi} ]/M^9.
\ea
The nonrelativistic fluid satisfies the continuity equation
$\dot{\rho}_m+3H \rho_m=0$.

We shall consider the case in which the late-time cosmic acceleration 
is realized by the field kinetic terms, i.e. $c_1=0$.
There is a de Sitter (dS) solution characterized by 
$H=H_{\rm dS}=$\,constant and 
$\dot{\phi}=\dot{\phi}_{\rm dS}=$\,constant. 
{}From Eqs.~(\ref{be1}) and (\ref{be2}) it then follows that 
\begin{eqnarray}
& & c_2 x_{\rm dS}^2=6+9\alpha-12\beta\,,
\label{ds1} \\
& & c_3 x_{\rm dS}^3=2+9\alpha-9\beta\,,
\label{ds2}
\end{eqnarray}
where $x_{\rm dS} \equiv \dot{\phi}_{\rm dS}/(H_{\rm dS}M_{\rm pl})$ and 
\be
\alpha \equiv c_4 x_{\rm dS}^4\,,\qquad
\beta \equiv c_5 x_{\rm dS}^5\,.
\ee
We normalize the mass $M$ to be $M^3=M_{\rm pl}H_{\rm dS}^2$.
Since $H_{\rm dS}$ is of the order of the present Hubble parameter
($H_0 \approx 10^{-60}M_{\rm pl}$), we have that 
$M \approx 10^{-40}M_{\rm pl}$.

In order to discuss the cosmological dynamics, it is convenient 
to introduce the following variables
\be
r_1 \equiv \frac{\dot{\phi}_{\rm dS}H_{\rm dS}}
{\dot{\phi}H}\,,\qquad
r_2 \equiv \frac{1}{r_1} \left( \frac{\dot{\phi}}
{\dot{\phi}_{\rm dS}} \right)^4\,.
\label{r1r2def}
\ee
At the dS point we have $r_1=1$ and $r_2=1$.
The Friedmann equation (\ref{be1}) can be written as 
$\Omega_{\rm DE}+\Omega_m=1$, where 
$\Omega_m \equiv \rho_m/(3M_{\rm pl}^2 H^2)$ and 
\be
\Omega_{\rm DE} \equiv \frac{\rho_{\rm DE}}{3M_{\rm pl}^2 H^2}
=-\frac12 (2+3\alpha-4\beta)r_1^3r_2+(2+9\alpha-9\beta)r_1^2r_2
-\frac{15}{2}\alpha r_1 r_2+7\beta r_2\,.
\label{Omede}
\ee
We also obtain the autonomous equations 
\ba
\label{eq:DRr1}
\hspace{-0.7cm}
r_1' &=& \frac{1}{\Delta} \left(r_1-1\right)
r_1 \left[ r_1 \left(r_1 (-3 \alpha +4 \beta -2)
+6 \alpha -5 \beta \right)-5 \beta \right] \nonumber\\
&&{}\times \left[ 18
+3 r_2 \left( r_1^3 (-3 \alpha +4\beta -2)+
2 r_1^2 (9 \alpha -9 \beta +2)-15 r_1 \alpha 
+14 \beta \right)\right]\,,\\
\label{eq:DRr2}
\hspace{-0.7cm}
r_2' &=& -\frac{1}{\Delta}
[ r_2 (6 r_1^2 (r_2 (45 \alpha ^2-4 (9 \alpha +2) \beta 
+36 \beta^2)+7 (9 \alpha -9 \beta +2))
+r_1^3 (-66 (3 \alpha -4 \beta +2) \nonumber \\ & &
-3 r_2 (-2 (201 \alpha +89) \beta +15\alpha
(9 \alpha +2)+356 \beta ^2))
-3 r_1 \alpha ( 123 r_2
\beta +36)+10 \beta ( 21 r_2 \beta -3) \nonumber \\ 
& &+3r_1^4 r_2 (9\alpha ^2-30 \alpha (4 \beta +1)+2 (2-9 \beta )^2) 
+3r_1^6 r_2 (3 \alpha -4 \beta+2)^2+3 r_1^5 r_2 
(9 \alpha -9 \beta +2) (3 \alpha -4 \beta +2))], 
\ea
where a prime represents a derivative with respect to $N=\ln a$, and 
\ba
\Delta &\equiv & 2 r_1^4 r_2 [ 72 \alpha ^2+30 \alpha  
(1-5 \beta )+(2-9 \beta )^2 ]+4 r_1^2 
[ 9 r_2 (5 \alpha ^2+9 \alpha \beta +(2-9 \beta ) \beta )
+2 (9 \alpha -9 \beta +2) ] \nonumber\\
&&+4 r_1^3 [ -3 r_2 \left(-2 (15 \alpha +1) \beta 
+3 \alpha  (9 \alpha +2)+4 \beta ^2\right)-3 \alpha 
+4 \beta -2]-24 r_1 \alpha  (16 r_2 \beta +3)
+10\beta (21 r_2 \beta +8)\,.
\ea
The Hubble parameter obeys the following equation
\be
\frac{H'}{H}=-\frac{5r_1'}{4r_1}-\frac{r_2'}{4r_2}\,.
\ee

We define the dark energy equation of state $w_{\rm DE}$ 
and the effective equation of state $w_{\rm eff}$, as 
\be
w_{\rm DE} \equiv \frac{p_{\rm DE}}{\rho_{\rm DE}}\,,
\qquad
w_{\rm eff} \equiv -1 -\frac{2H'}{3H}\,.
\ee
Using the continuity equation 
$\dot{\rho}_{\rm DE}+3H(\rho_{\rm DE}+p_{\rm DE})=0$, 
we obtain the relation $w_{\rm DE}=
w_{\rm eff}-\Omega_{\rm DE}'/(3\Omega_{\rm DE})$.

There are three distinct fixed points: (A) $(r_1, r_2)=(0, 0)$, 
(B) $(r_1, r_2)=(1, 0)$, and (C) $(r_1, r_2)=(1, 1)$.
As we see from the definition in Eq.~(\ref{r1r2def})
the point (C) corresponds to the dS solution, which
is always classically stable \cite{DT3}.
The point (B) is a tracker solution found in Ref.~\cite{DT2}, 
along which the field velocity evolves as $\dot{\phi} \propto 1/H$.
The fixed point (B) is followed by the stable dS solution
once $r_2$ grows to the order of 1.
Depending on the initial conditions of $r_1$, the epoch at which 
the solutions approach the tracker is different.

In the following we summarize the background evolution 
in two regimes: (i) $r_1 \ll 1, r_2 \ll 1$ and 
(ii) $r_1=1$ \cite{DT2,DT3}.
\begin{itemize}
\item (i) $r_1 \ll 1, r_2 \ll 1$

This is the regime in which the term ${\cal L}_5$ gives the
dominant contribution in Eq.~(\ref{Omede}), i.e.
\be
\Omega_{\rm DE} \simeq 7 \beta r_2\,.
\ee
Since $r_1' \simeq 9r_1/8$ and $r_2' \simeq 3r_2/8$ from 
Eqs.~(\ref{eq:DRr1}) and (\ref{eq:DRr2}), the two variables 
$r_1$ and $r_2$ evolve as 
\be
r_1 \propto a^{9/8}\,,\qquad 
r_2 \propto a^{3/8}\,.
\ee
During the matter era we have 
\be
w_{\rm DE} \simeq -1/8\,,\qquad
w_{\rm eff} \simeq 0\,.
\ee
If $\beta r_2>0$, then the scalar ghosts do not appear \cite{DT2,DT3}.
For the initial conditions with $r_2>0$ we require that 
\be
\beta>0\,.
\ee
The conditions for the avoidance of Laplacian instabilities 
are automatically satisfied \cite{DT2,DT3}.

\item (ii) $r_1=1$

Along the tracker characterized by $r_1=1$, there is 
a simple relation 
\be
\Omega_{\rm DE}=r_2\,.
\ee
In the regime $r_2 \ll 1$ one has
$r_2' \simeq 6r_2$ from Eq.~(\ref{eq:DRr2}), 
so that the evolution of $r_2$ during the deep matter 
era is given by 
\be
r_2 \propto a^6\,.
\label{r2tracker}
\ee
Along the tracker the evolution of the Hubble parameter can be
analytically known as \cite{NFT}
\be
\left( \frac{H(z)}{H_0} \right)^2=\frac12 \Omega_{m0}
(1+z)^3+\sqrt{1-\Omega_{m0}+\frac{(1+z)^6}{4}
(\Omega_{m0})^2}\,,
\label{trackerpara}
\ee
where $z=a_0/a-1$ is the redshift ($a_0$ is the scale factor today), 
and $\Omega_{m0}$ is the density parameter of nonrelativistic
matter today.
Even in the presence of radiation and the cosmic 
curvature there exists an analytic form of $H(z)$ \cite{NFT}.

Since $\rho_{\rm DE}=3M^6/H^2$ and $p_{\rm DE}=
-3M^6 (2+w_{\rm eff})/H^2$, it follows that 
\be
w_{\rm DE}=-2-w_{\rm eff}=-\frac{2}{r_2+1}\,,\qquad
w_{\rm eff}=-\frac{2r_2}{r_2+1}\,.
\ee
In the deep matter era ($r_2 \ll 1$) we have $w_{\rm DE}=-2$ and 
$w_{\rm eff}=0$, whereas at the dS solution 
$w_{\rm DE}=w_{\rm eff}=-1$.
The conditions for the avoidance of ghosts and Laplacian 
instabilities have been derived in Refs.~\cite{DT2,DT3}. 
The allowed parameter space in the $(\alpha,\beta)$ plane
is summarized in figure 1 of Ref.~\cite{DT2}.
\end{itemize}

It should be pointed out that the tracker solution and the ${\cal L}_5$-dominant solution can be distinguished also by means of the following physical quantity. 
When $c_1=0$ one has $\partial L/\partial\phi=0$, where $L \equiv \sqrt{-g}\mathcal{L}$
is the Lagrangian including the volume factor $\sqrt{-g}$ in Eq.~(\ref{action}).
Then the conjugate momentum to the field $\phi$ is conserved, so that
$\partial L/\partial \dot\phi={\rm constant}$. This constraint can be written as
\begin{equation}
\label{eq:constraint}
P_\phi\equiv{\frac {{r_2}^{1/4}\, a ^{3}
 \left(1- r_1 \right)  \left( 2\,{r_1}^{2}+3\alpha\,{r_1}^{2}
-4\beta\,{r_1}^{2}-6\,\alpha\,r_1+5\beta\,r_1+5
\,\beta \right) }{{r_1}^{11/4}}}={\rm constant}\, .
\end{equation}
On the tracker $P_\phi$ exactly vanishes\footnote{The tracker can then be seen 
as the zero momentum solutions.}, whereas for the ${\cal L}_5$-dominant solution
$P_\phi \simeq 5\beta\,a^3\,(r_2/r_1^{11})^{1/4}\neq 0$. 
Although Eq.~(\ref{eq:constraint}) is not independent of the others, 
we will use it later on in order to check the good convergence 
of numerical solutions.

\section{Cosmological perturbations}
\label{seccosmoper}

Let us consider scalar metric perturbations $\Psi$, $\Phi$, and 
$\chi$ about the flat FLRW background \cite{Bardeen}:
\be
{\rm d}s^{2}=-(1+2\Psi){\rm d}t^{2}-2\partial_{i}
\chi {\rm d}t {\rm d}x^{i}
+a^2(t)(1+2\Phi)\delta_{ij} {\rm d}x^{i} {\rm d}x^{j}\,,
\label{permetric}
\ee
where we have chosen the spatial gauge such that the spatial metric is
diagonal\footnote{In the published PRD version there is an extra factor $a(t)$ in front of 
the $\partial_i \chi$ term. We corrected this typo. }. 
There is still a freedom to fix the temporal part $\xi_{0}$
of a vector associated with a scalar gauge transformation.  Later on
we will discuss this gauge degree of freedom in more detail.

The field $\phi$ is decomposed into the background and inhomogeneous
parts, as $\phi (t, {\bm x})= \tilde{\phi}(t)+\delta \phi (t, {\bm x})$.
In the following we omit the tilde for the background quantity. 
We shall use the action approach rather than taking 
the usual step of linearizing the equations of motion, 
as the former is more convenient from a computational point of view 
(indeed the scalar Lagrangian is quite compact compared to the 
tensorial equations of motion). 
Evidently the two approaches lead to equivalent equations of motion with 
the same set of solutions.  
For the perfect fluid, we cannot set $w=0$ in Eq.~(\ref{action}) identically from 
the beginning as its action would vanish. However, we can consider a fluid 
whose equation of state parameter $w$ is different from zero, 
and only after we obtain the equations of motion for the perturbation variables, we will take the limit $w \to 0$. 
Since we are interested in the scalar perturbations of a cold fluid 
(temperature $T=0$) with a barotropic equation of state of the form
$p_m = w \rho_m$ (in the limit $w \to 0$), the action approach of linear 
perturbation theory for a perfect fluid introduced in Refs.~\cite{actPF} 
can be simplified as follows. 

Let us define a 4-velocity of the perfect fluid,
as $u_\alpha=\mu^{-1}\,\partial_\alpha {\ell}$. 
This relation defines $\ell$, which completely describes the scalar component 
of $u_\alpha$ for the case under study.
The normalization of the 4-velocity ($u_{\alpha} u^{\alpha}=-1$) implies 
that $\mu=\sqrt{-g^{\alpha \beta}\partial_\alpha {\ell}\partial_\beta {\ell}}$. 
Then at linear level we have $\delta\mu=-\dot{\delta\ell}-\mu\Psi$.
Now the quantity $\sqrt{-g}\, p_m(\mu,s)$ can be expanded at second order 
in the gravitational fields and in $\delta\ell$ and $\delta s$. 
However, since the fluid is cold ($T=0$), the first principle of thermodynamics, 
${\rm d}p_m=n{\rm d}\mu-nT{\rm d}s$, imposes that 
$(\partial p_m/\partial s)_\mu$,
$(\partial^2 p_m/\partial s^2)_\mu$, and 
$(\partial^2 p_m/\partial s\partial\mu)$ 
vanish\footnote{This also implies that $n=n(\mu)$. 
For example, $\mu\propto n^{w}$ is enough to give $p_m=w\rho_m$.}.
In turn this implies that $p_m$ depends on $\mu$ alone, and we do not need to 
expand the action in terms of $\delta s$, the entropy perturbation 
field\footnote{In the equations of motion approach, this result corresponds to the fact that, 
once $p_m=w\rho_m$, 
no entropy perturbations appear in the dynamical equations.}. 
Since the action for matter is a function of $\mu$ alone, as $p_m=p_m(\mu)$, 
and $\delta\mu$ is a function of the perturbation variables 
$\delta\ell$ and $\Psi$, the barotropic perfect fluid introduces 
only one new independent perturbed scalar field, $\delta\ell$.
Furthermore, since $\rho_m \equiv n\mu-p_m$, it follows that 
${\rm d}\rho_m=\mu\,{\rm d} n+nT{\rm d} s=
\mu\,{\rm d} n=\mu\dot n\,\delta\mu/\dot\mu$
at linear level. 
Using the field redefinition $\delta\ell=-\mu v$, where 
$v$ is a velocity potential for the perfect fluid, after some algebra, 
we find 
\be
\dot{v}-3Hwv=\Psi+\frac{w}{1+w}\delta\,,
\label{delrhoeq2}
\ee
where $\delta \equiv \delta \rho_m/\rho_m$ is 
the density contrast. 

In Fourier space the density contrast obeys the following 
equation of motion \cite{Hwang}
\begin{equation}
\dot{\delta}=-(1+w) \left( 3\dot{\Phi}+
\frac{k^2}{a^2}v-\frac{k^2}{a^2}\,\chi \right)\,,
\label{delrhoeq1} \\
\end{equation}
where $k$ is a comoving wave number.
This equation follows from the first-order part of the 
continuity equation $\nabla_{\mu} T^{\mu}_{0}=0$.
In the following we derive other perturbation equations from 
the second-ord8er action for perturbations. 
In Appendix \ref{sec:bianchi} we will show how to obtain
Eq.~(\ref{delrhoeq1}) from the action approach.

Let us consider nonrelativistic matter with $w=0$.
We introduce the gauge-invariant matter perturbation 
$\delta_m$, as \cite{Bardeen}
\be
\delta_m \equiv \delta+3Hv\,.
\label{delmdef}
\ee
Using Eqs.~(\ref{delrhoeq1}) and (\ref{delrhoeq2}), it follows that
\be
\ddot{\delta}_m+2H\dot{\delta}_m+\frac{k^2}{a^2} (\Psi - \dot{\chi})
=3(\ddot{Q}+2H \dot{Q})\,,
\label{mattereq}
\ee
where $Q \equiv Hv-\Phi$.

Let us now derive the coupled equations for metric/field/matter
perturbations. Expansion of the action (\ref{action}) 
at second order gives
\be
[\sqrt{-g}\mathcal{L}]^{(2)}\equiv L(\Psi,\Phi,\chi,\delta\phi,v)\,.
\label{eq:2ndact}
\ee
Equation (\ref{delrhoeq2}) shows that $\delta$ and $\Psi$ 
are not independent fields. 
In the second-order action (\ref{eq:2ndact}) 
we have chosen the latter field.
If we vary the Lagrangian (\ref{eq:2ndact}) with respect to $\Psi$, 
then we find a term $\rho_m (\Psi-\dot{v})/w$
that leads to an apparent singularity for $w \to 0$.
However, Eq.~(\ref{delrhoeq2}) allows to reduce this term 
to $-\rho_m [\delta/(1+w)+3Hv]$.
Taking the limit $w \to 0$ at the end, 
we obtain the equation of motion
\be
E_{\Psi} \equiv 
A_1 \dot{\Phi}+A_2 \dot{\delta \phi}-\rho_m \dot{v}
+A_3 \frac{k^2}{a^2}\Phi+A_4\Psi
+A_5 \frac{k^2}{a^2}\chi
+\left( \frac{1}{2}c_1 M^3+A_6 \frac{k^2}{a^2} \right) 
\delta \phi-\rho_m \delta=0\,,
\label{eq:Psi}
\ee
where the coefficients $A_i$ are given 
in Appendix \ref{coefficients}. 
For the theories up to the Lagrangian ${\cal L}_2$ we have checked that, after substituting 
the relation $\dot{v}=\Psi$ [which comes from Eq.~(\ref{delrhoeq2})] and 
using the background Friedmann equation for the coefficient of $\Psi$, 
Eq.~(\ref{eq:Psi}) reduces to the usual 0-0 component of the perturbed Einstein equation
obtained in the literature (see e.g. ~\cite{mabert}).

Variations of the second-order Lagrangian $L$ with respect to 
$\Phi$, $\chi$, and $\delta \phi$ result in the following equations
\ba
& &\label{eq:Phi}
E_{\Phi} \equiv 
B_1 \ddot{\Phi}+B_2 \ddot{\delta \phi}+B_3 \dot{\Phi}
+B_4 \dot{\delta \phi}+B_5 \dot{\Psi}+B_6 \frac{k^2}{a^2}\Phi
+\left( B_7 \frac{k^2}{a^2}+\frac32 c_1M^3 \right) 
\delta \phi \nonumber \\
& &~~~~~~~~+ \left( B_8 \frac{k^2}{a^2}+B_9 \right)\Psi
+B_{10} \frac{k^2}{a^2} \dot{\chi}
+B_{11} \frac{k^2}{a^2} \chi+3 \rho_m \dot{v}=0\,,\\
& &\label{eq:chi}
E_{\chi} \equiv 
C_1 \dot{\Phi}+C_2 \dot{\delta \phi}+C_3 \Psi
+C_4 \delta \phi+\rho_m v=0\,,\\
& &\label{eq:delphi}
E_{\delta \phi} \equiv 
D_1 \ddot{\Phi}+D_2 \ddot{\delta \phi}+D_3 \dot{\Phi}
+D_4 \dot{\delta \phi}+D_5 \dot{\Psi}+D_6 \frac{k^2}{a^2}
\dot{\chi} \nonumber \\
& &~~~~~~~~~+\left( D_7 \frac{k^2}{a^2}+D_8 \right)\Phi
+D_9 \frac{k^2}{a^2} \delta \phi+\left( D_{10} \frac{k^2}{a^2}
+D_{11} \right) \Psi+D_{12} \frac{k^2}{a^2}\chi=0\,,
\ea
where the coefficients $B_i$, $C_i$, and $D_i$ are 
given in Appendix \ref{coefficients}. 
$E_{\Phi}$ corresponds to the spatial trace of the perturbed 
modified Einstein equations, $E_\chi$ to the momentum constraint, 
and $E_{\delta \phi}$ to the 
perturbed equation of motion for the Galileon field.
Varying the Lagrangian $L$ in terms of $v$ 
gives rise to Eq.~(\ref{delrhoeq2}), i.e.
$\dot{v}=\Psi$ for $w=0$.

Another constraint equation can be found by introducing the perturbation 
$\gamma$ in the metric (\ref{permetric}), 
as $\delta g_{ij}=2a^2\partial_i\partial_j\gamma$. 
Its equation of motion is given by 
\be
E_{\gamma} \equiv
\frac13 (B_1 \ddot{\Phi}+B_2 \ddot{\delta \phi}+B_3 \dot{\Phi}
+B_4 \dot{\delta \phi}+B_5 \dot{\Psi}+B_9 \Psi)
+\frac12 c_1 M^3 \delta \phi+\rho_m \dot{v}=0\,,
\ee
which does not contain $\gamma$ explicitly (as the quadratic terms 
in $\gamma$ in the action can be integrated out).
Evaluating $E_{\Phi}-3E_\gamma$ 
leads to the following equation
\begin{equation}
\label{eq:constr2}
\tilde{E}_{\gamma} \equiv B_{6}\,\Phi+B_{7}\,\delta\phi
+B_{8}\,\Psi+B_{10}\,\dot\chi+B_{11}\,\chi=0\,.
\end{equation}
The same relation can be derived by using the Bianchi identities, 
see Appendix \ref{sec:bianchi}.

In what follows we choose the longitudinal gauge with $\chi=0$. 
This choice, together with imposing $\gamma=0$, completely fixes the gauge.
We also set $c_1=0$ to discuss the evolution of 
cosmological perturbations.

\section{Quasistatic approximation on subhorizon scales}
\label{secsubho}

Let us consider the evolution of perturbations for the modes deep 
inside the Hubble radius.
We derive the equations of matter perturbations and gravitational 
potentials under the quasistatic approximation on subhorizon scales
($k \gg aH$). This corresponds to the approximation under which
the dominant contributions to the perturbation equations are those
including $k^2/a^2$ and $\delta$ \cite{quasi1,quasi2}.
In the longitudinal gauge we obtain the following approximate 
equations from Eqs.~(\ref{eq:Psi}), (\ref{eq:constr2}), and (\ref{eq:delphi}):
\ba
& &\label{ap1}
\frac{k^2}{a^2} \left( A_3 \Phi+A_6 \delta \phi \right) \simeq 
\rho_m \delta\,,\\
& &\label{ap2} 
B_6 \Phi+B_7 \delta \phi+B_8 \Psi = 0\,,\\
& & \label{ap3} 
D_7 \Phi+D_9 \delta \phi+D_{10}\Psi \simeq 0\,. 
\ea

{}From Eqs.~(\ref{ap2}) and (\ref{ap3}) one can express 
$\Phi$ and $\delta \phi$ in terms of $\Psi$, i.e.
\ba
\Phi &\simeq& \frac{A_3 D_9-A_6 B_7}{B_7^2 -B_6 D_9}\Psi\,,
\label{apper1}\\
\delta \phi &\simeq&
\frac{A_6 B_6-A_3 B_7}{B_7^2 -B_6 D_9}\Psi
\label{apper2}\,,
\ea
where we have used $B_8=A_3$, $D_7=B_7$, and $D_{10}=A_6$.
Substituting Eqs.~(\ref{apper1}) and (\ref{apper2}) into Eq.~(\ref{ap1}), 
it follows that 
\be
\frac{k^{2}}{a^{2}}\Psi \simeq 
-4\pi G_{{\rm eff}}\rho_m \delta\,,
\label{eq:Lapla}
\ee
where
\be
G_{\rm eff} \equiv \frac{2M_{\rm pl}^2(B_7^2-B_6 D_9)}
{2A_3A_6B_7-A_3^2 D_9-A_6^2 B_6}G\,.
\ee
Since $|\dot{\delta}|$ is at most of the order of $|H \delta|$ 
in Eq.~(\ref{delrhoeq1}), 
one has $|(k^2/a^2)v| \lesssim |H \delta|$.
This means that $|Hv/\delta| \lesssim (aH/k)^2 \ll 1$ for subhorizon modes, 
which leads to $\delta_m \simeq \delta$ in Eq.~(\ref{delmdef}).
Since the r.h.s. of Eq.~(\ref{mattereq}) can be neglected relative to the 
l.h.s., we obtain the equation for the gauge-invariant matter perturbation 
for the modes deep inside the Hubble radius:
\be
\ddot{\delta}_m+2H\dot{\delta}_m
-4\pi G_{\rm eff} \rho_m \delta_m \simeq 0\,,
\label{mattereqap}
\ee
where we have used the Poisson equation (\ref{eq:Lapla}).

{}From Eqs.~(\ref{apper2}) and (\ref{eq:Lapla}) the field perturbation 
is given by 
\be
\delta \phi \simeq \frac{3M_{\rm pl}^2(A_3 B_7-A_6 B_6)}
{2A_3A_6 B_7 -A_3^2 D_9-A_6^2 B_6}
\left( \frac{aH}{k} \right)^2 \Omega_m \delta\,.
\label{delap}
\ee
We introduce the following quantity to describe the difference 
between the two gravitational potentials:
\be
\eta \equiv -\frac{\Phi}{\Psi}
 \simeq \frac{A_3 D_9-A_6 B_7}{B_6 D_9 -B_7^2}\,,
 \label{etadef}
\ee
where we have used Eq.~(\ref{apper1}) in the last approximate 
equality.
We also define the effective gravitational potential 
\be
\Phi_{\rm eff} \equiv (\Psi-\Phi)/2\,,
\ee
which is related with the deviation of the light rays
in CMB and weak lensing observations \cite{obsermo}.
{}From the Poisson equation (\ref{eq:Lapla}) it follows that 
\be
\frac{k^2}{a^2}\Phi_{\rm eff} \simeq
-4\pi G_{\rm eff} \frac{1+\eta}{2} \rho_m \delta\,.
\ee
Using the density parameter $\Omega_m=8 \pi G \rho_m/(3H^2)$
and the relation $\delta_m \simeq \delta$, we have 
\be
\Phi_{\rm eff} \simeq -\frac32 \frac{G_{\rm eff}}{G}
\frac{1+\eta}{2} \Omega_m \delta_m 
\left( \frac{aH}{k} \right)^2\,.
\label{Phieffap}
\ee

The matter perturbation equation (\ref{mattereqap}) can be 
written as 
\be
\delta_m''+\left( 2+\frac{H'}{H} \right) \delta_m'
-\frac32 \frac{G_{\rm eff}}{G} \Omega_m \delta_m
\simeq 0\,.
\label{delmeqd}
\ee
In the limit that $\dot{\phi} \to 0$ and $\ddot{\phi} \to 0$, 
the effective gravitational coupling $G_{\rm eff}$
reduces to $G$. Hence, in the early cosmological epoch, 
the General Relativistic behavior is recovered.
If $G_{\rm eff} \simeq G$, then the evolution of matter 
perturbations during the deep matter era 
($H'/H \simeq -3/2$ and $\Omega_m \simeq 1$)
is given by $\delta_m \propto a \propto t^{2/3}$.

As the field velocity grows in time, $G_{\rm eff}$
is subject to change. This leads to the modified growth 
rate of matter perturbations compared to the $\Lambda$-cold dark matter 
($\Lambda$CDM) model.
Unlike $f(R)$ gravity \cite{quasi2} and Brans-Dicke 
theory \cite{Bransper} the effective 
gravitational coupling is independent of the wave number 
$k$. The quantity $\eta$ defined in Eq.~(\ref{etadef})
reduces to 1 for $\dot{\phi} \to 0$ and $\ddot{\phi} \to 0$, 
but it deviates from 1 with the growth of 
$\dot{\phi}$ and $\ddot{\phi}$.
In three different regimes characterized by 
(i) $r_1 \ll 1$, $r_2 \ll 1$, 
(ii) $r_1=1$, $r_2 \ll 1$, and 
(iii) $r_1=1$, $r_2=1$, 
one can estimate $G_{\rm eff}$ and $\eta$ as follows.
\begin{itemize}
\item (i) $r_1 \ll 1$, $r_2 \ll 1$

In this case we expand $G_{\rm eff}$ and $\eta$ about $r_1=0, r_2=0$.
Together with the use of the background equations (\ref{eq:DRr1}) 
and (\ref{eq:DRr2}), we obtain
\ba
\label{Geffap}
\frac{G_{\rm eff}}{G} &=& 1+\left( \frac{255}{8}\beta+\frac{211}{16}
\alpha r_1 \right) r_2+{\cal O} (r_2^2)\,,\\
\label{etaap}
\eta &=& 1+\left( \frac{129}{8}\beta+\frac{589}{16}
\alpha r_1 \right) r_2+{\cal O} (r_2^2)\,.
\ea
In this epoch the cosmological dynamics is dominated 
by the term ${\cal L}_5$.
Since $\beta>0$ to avoid ghosts, we have that 
$G_{\rm eff}>G$ and $\eta>1$ in this regime.
Hence the growth rates of $\delta_m$ and $\Phi_{\rm eff}$ 
are larger than those in the $\Lambda$CDM model.

\item (ii) $r_1=1$, $r_2 \ll 1$

Expanding $G_{\rm eff}$ and $\eta$ about $r_2=0$, it follows that 
\ba
\frac{G_{\rm eff}}{G} &=& 1+\frac{291\alpha^2+702\beta^2
-933\alpha \beta+20\alpha-84\beta+4}
{2(10\alpha-9\beta+8)}r_2+{\cal O}(r_2^2) \,,
\label{Gefftracker}\\
\eta &=& 1-\frac{3(126\alpha^2+306 \beta^2-405\alpha \beta
+4\alpha-30\beta)}{2(10\alpha-9\beta+8)}r_2+{\cal O}(r_2^2) \,.
\label{etatracker}
\ea
The evolution of $G_{\rm eff}$ and $\eta$ depends on both 
$\alpha$ and $\beta$.
If $\alpha=1.4$ and $\beta=0.4$, for example, 
we have $G_{\rm eff}/G \simeq 1+4.31r_2$ and 
$\eta \simeq 1-5.11r_2$, respectively.
In this case $G_{\rm eff}>G$, but $\eta$ is smaller 
than 1.

\item (iii) $r_1=1$, $r_2=1$

At the dS point we have
\ba
\label{Geffde}
\frac{G_{\rm eff}}{G} &=& \frac{1}{3(\alpha-2\beta)} \,,\\
\label{etade}
\eta &=& 1 \,.
\ea
This means that $\eta$ is not subject to change compared to 
the $\Lambda$CDM model.
In Refs.~\cite{DT2,DT3} it was shown that 
the viable parameter space consistent with 
the absence of ghosts and instabilities is confined 
in the region $2\beta \le \alpha \le 2\beta+2/3$.
{}From Eq.~(\ref{Geffde}) the effective gravitational 
coupling is constrained to be $G_{\rm eff}/G \ge 1/2$.
On the line $\alpha=2\beta$, $G_{\rm eff}$ 
goes to infinity (this includes 
the case in which only the terms up to ${\cal L}_3$
are present, i.e. $\alpha=\beta=0$).
On the line $\alpha-2\beta=1/3$, the effective 
gravitational coupling is equivalent to $G$. 
 
\end{itemize}

For the Lagrangian up to ${\cal L}_3$ one has $\eta=1$, but 
the presence of scalar nonminimal couplings with curvature terms
in ${\cal L}_4$ and ${\cal L}_5$ 
(such as $R (\nabla \phi)^2/2$) gives rise to 
the situation where $\eta$ is different from 1.
The above approximate formulas are useful to discuss 
the evolution of perturbations on the scales relevant to
large-scale structures.

\section{Numerical results}
\label{secnume}

In this section we present numerical results for the evolution 
of perturbations without employing the quasistatic approximation 
on subhorizon scales.
The accuracy of the quasistatic approximation will be 
confirmed for the modes $k \gg aH$.

Among the six equations (\ref{eq:Psi})-(\ref{eq:constr2}), 
three of them are independent.
For the numerical purpose we solve Eqs.~(\ref{eq:Psi}), (\ref{eq:chi}), 
and (\ref{eq:constr2}) together with Eqs.~(\ref{delrhoeq2}) 
and (\ref{delrhoeq1}). 
We have also confirmed that $P_{\phi}$ defined in Eq.~(\ref{eq:constraint})
remains constant up to the accuracy of $10^{-8}$ during the whole evolution.
It is convenient to introduce the following dimensionless variables
\be
V \equiv Hv\,,\qquad \delta \varphi \equiv 
\delta \phi/(x_{\rm dS}M_{\rm pl})\,,
\ee
with $\tilde{A}_1 \equiv A_1/(HM_{\rm pl}^2)$, 
$\tilde{A}_2 \equiv x_{\rm dS}A_2/(HM_{\rm pl})$, 
$\tilde{A}_3 \equiv A_3/M_{\rm pl}^2$,
$\tilde{A}_4 \equiv A_4/(H^2M_{\rm pl}^2)$,
$\tilde{A}_6 \equiv x_{\rm dS}A_6/M_{\rm pl}$, 
$\tilde{B}_6 \equiv B_6/M_{\rm pl}^2$, 
$\tilde{B}_7 \equiv x_{\rm dS}B_7/M_{\rm pl}$, and 
$\tilde{C}_4 \equiv x_{\rm dS}C_4/(HM_{\rm pl})$. 
In the longitudinal gauge we obtain the following 
equations from Eqs.~(\ref{eq:Psi}), (\ref{eq:chi}), 
(\ref{eq:constr2}), (\ref{delrhoeq1}), 
and (\ref{delrhoeq2}):
\ba
& & \Psi=-(\tB_6 \Phi+\tB_7 \delta \varphi)/A_3\,,
\label{per1}\\
& & \Phi'= [( 3\tA_4 \tA_6 \tB_6+\tA_1 \tA_2 \tB_6 
-3\tA_3^2 \tA_6 k^2/(aH)^2-9 \tA_6 \tB_6 \Omega_m )\Phi 
\nonumber \\
& &~~~~~~~
+ ( 3\tA_2 \tA_3 \tC_4+3\tA_4 \tA_6 \tB_7
+\tA_1 \tA_2 \tB_7-9\tA_6 \tB_7 \Omega_m-3\tA_3 \tA_6^2
k^2/(aH)^2 )\delta \varphi+9\tA_3 \tA_6 \Omega_m \delta+
9\tA_2 \tA_3 \Omega_m V] \nonumber \\
& &~~~~~~~\times
[3\tA_3 (\tA_1 \tA_6 -\tA_2 \tA_3)]^{-1}\,,
\label{per2} \\
& & \delta \varphi'=-[(\tA_1^2 \tB_6+3\tA_3 \tA_4 \tB_6
-3\tA_3^3 k^2/(aH)^2-9 \tA_3 \tB_6 \Omega_m )\Phi
\nonumber \\
& &~~~~~~~~
+ (\tA_1^2 \tB_7+3\tA_3 \tA_4 \tB_7+3\tA_1 \tA_3 \tC_4
-3 \tA_3^2\tA_6 k^2/(aH)^2-9\tA_3 \tB_7 \Omega_m)
\delta \varphi+9 \tA_3^2
\Omega_m\delta+9\tA_1 \tA_3 \Omega_m V ]\nonumber \\
& &~~~~~~~~\times 
[3\tA_3 (\tA_1 \tA_6 -\tA_2 \tA_3)]^{-1}\,,
\label{per3} 
\\
& & \delta'=-3\Phi'-k^2/(aH)^2\,V\,,
\label{per4} 
\\
& & V'=(H'/H)V+\Psi\,,
\label{per5}
\ea
where we have used $\tB_8=\tA_3$, $\tC_1=\tA_3$, 
$\tC_2=\tA_6$, and $\tC_3=-\tA_1/3$.
The time-dependent coefficients $\tA_1$, $\tB_6$,
e.t.c. can be expressed by using the variables 
$\alpha$, $\beta$, $r_1$, and $r_2$.
Solving the perturbation equations (\ref{per1})-(\ref{per5})
together with the background equations (\ref{eq:DRr1}) 
and (\ref{eq:DRr2}), we find the evolution of 
$\Psi$, $\Phi$, $\delta \varphi$, $\delta$, and $V$.

During the deep matter era in which the field perturbation is 
negligibly small, the evolution of $\Psi$, $\Phi$, $\delta$, 
and $V$ is similar to that in GR.
The initial conditions are chosen to satisfy 
$\Phi'=0$ and $\delta \varphi'=0$.
{}From Eqs.~(\ref{per2}) and (\ref{per3}) this gives 
two constraints on the variables $\Phi$, $V$, 
$\delta \varphi$, and $\delta$.
For given $\delta \varphi$ and $\delta$, the initial conditions
of $\Phi$ and $V$ are determined accordingly.
For the modes deep inside the Hubble radius, 
Eq.~(\ref{delap}) provides a relation between 
$\delta \varphi$ and $\delta$.
We choose the initial condition $\delta=10^{-5}$, but 
this can be chosen to be any value (as long as 
the amplitudes of perturbations do not matter). 
We note that the relation (\ref{delap})
cannot be used for the wavelengths larger than
the Hubble radius.
Apart from the modes deep inside
the Hubble radius at the onset of integration, 
we adopt the initial condition $\delta \varphi=0$.
Provided that $\delta \varphi$ is small initially, 
the dynamics of perturbations is similar to
that in the case $\delta \varphi=0$. 
For large initial values of $\delta \varphi$, 
the field perturbation tends to oscillate for
small-scales modes (associated with large
wave numbers $k$). 
This situation is similar to what happens in the 
generalized Galileon model \cite{Kobayashi1}.

\begin{figure}
\begin{centering}
\includegraphics[width=3.2in,height=3.5in]{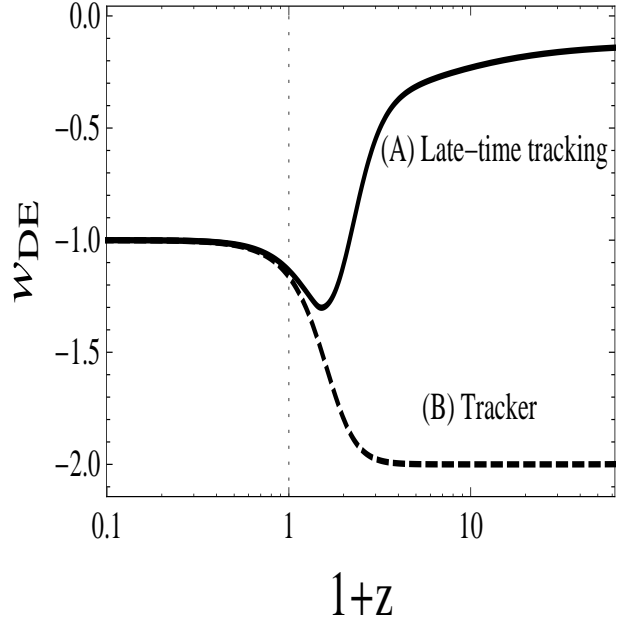} 
\par\end{centering}
\caption{The equation of state $w_{\rm DE}$ versus 
the redshift $z$ for the model parameters 
$\alpha=1.37$ and $\beta=0.44$
with two different initial conditions:
(A) $r_1=0.03$, $r_2=0.003$, and 
(B) $r_1=0.999$, $r_2=7.0 \times 10^{-11}$.
The case (B) corresponds to the tracker solution, whereas
in the case (A) the solution approaches the tracker 
around today. The present epoch ($z=0$) is shown as 
a dotted line (we also draw the dotted line in other figures).}
\centering{}\label{wde} 
\end{figure}

We identify the present epoch (the redshift $z=0$) to be
$\Omega_{\rm DE}=0.72$. 
In order to find the evolution of the quantity 
$k/(aH)$, we also integrate the following equation
\be
b'=b \left(1+H'/H \right)\,,
\ee
where $b \equiv aH$. 
The scales relevant to the linear regime 
of the galaxy power spectrum are
$30 \lesssim k/(a_0H_0) \lesssim 600$, 
where the subscript ``0'' represents present values.
Note that the above upper limit corresponds to 
$k=0.2\,h$\,Mpc$^{-1}$, where 
$h=0.72 \pm 0.08$ \cite{Percival}.
For the scales $k/(a_0H_0) \gtrsim 600$ the 
nonlinear effect becomes crucially important.
Meanwhile, the wave numbers relevant to the 
ISW effect in CMB anisotropies 
correspond to the large-scale modes with 
$k/(a_0H_0)={\cal O}(1)$.

In Ref.~\cite{NFT} two of the present authors placed observational 
constraints on the Galileon model (\ref{action}) from 
the background expansion history of the Universe.
The combined data analysis of the type Ia 
supernovae (Constitution and Union2 sets), the CMB
shift parameters (WMAP7), and the baryon acoustic oscillations (BAO)
show that the tracker solution described by 
Eq.~(\ref{trackerpara}) is not favored, but the solutions 
that approach the tracker at late times can be compatible 
with the background observational constraints.
As illustrated in Fig.~\ref{wde}, 
the equation of state $w_{\rm DE}$ 
for the tracker [case (B)] evolves from $-2$ (deep matter era) to $-1$
(dS epoch). The solution (A) in Fig.~\ref{wde}
starts to evolve from the value $w_{\rm DE}=-1/8$ and then it 
enters the tracking regime around the present epoch.
In the flat FLRW background the best-fit model 
parameters are
$\alpha=1.411 \pm 0.056$, $\beta=0.422 \pm 0.022$
(Constitution$\ +\ $CMB$\ +\ $BAO, 68\,\% C.L.), and 
$\alpha=1.404 \pm 0.057$, $\beta=0.419 \pm 0.023$
(Union2$\ +\ $CMB$\ +\ $BAO, 68\,\% C.L.).

\begin{figure}
\begin{centering}
\includegraphics[width=3.3in,height=3.2in]{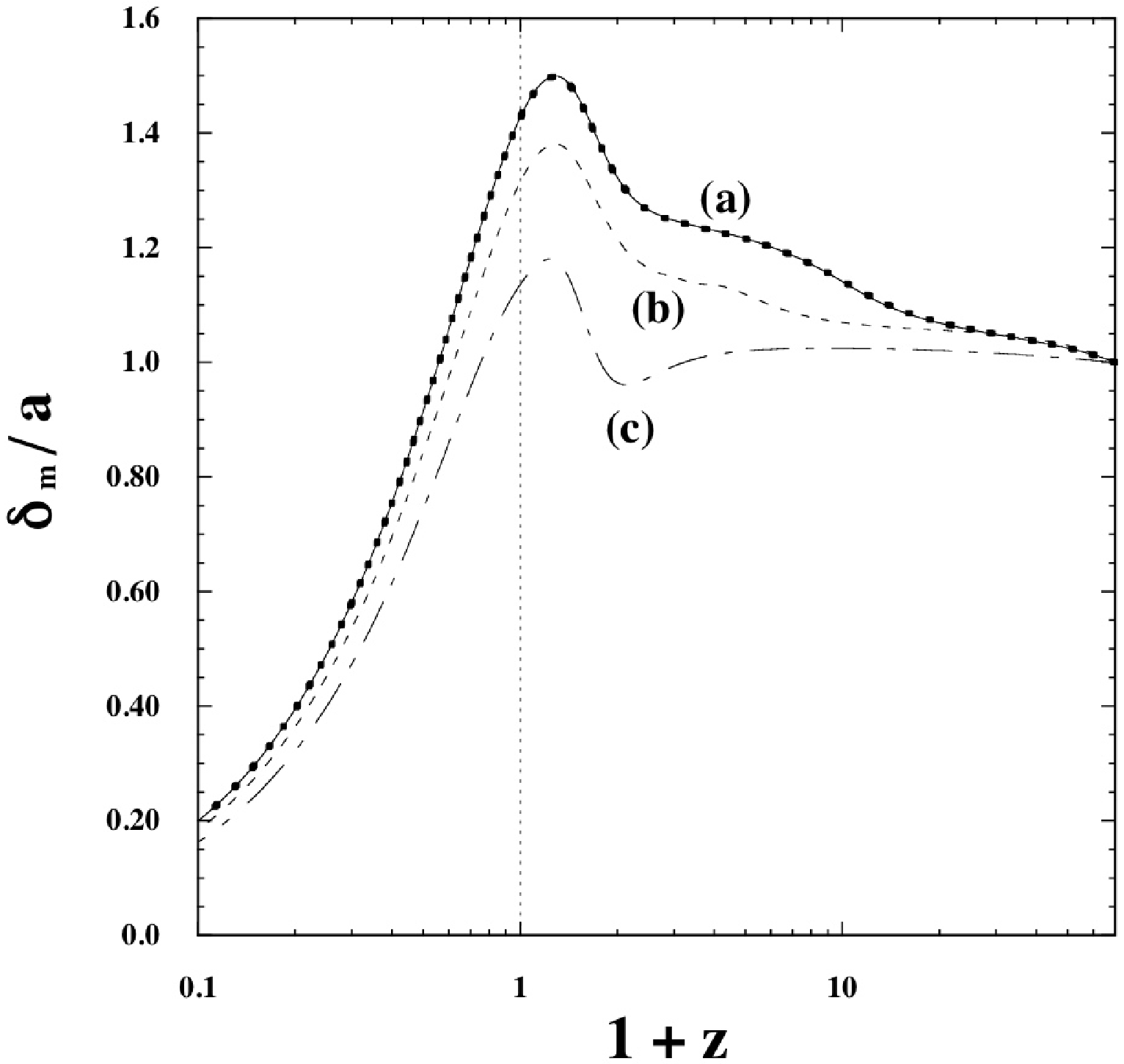} 
\includegraphics[width=3.3in,height=3.2in]{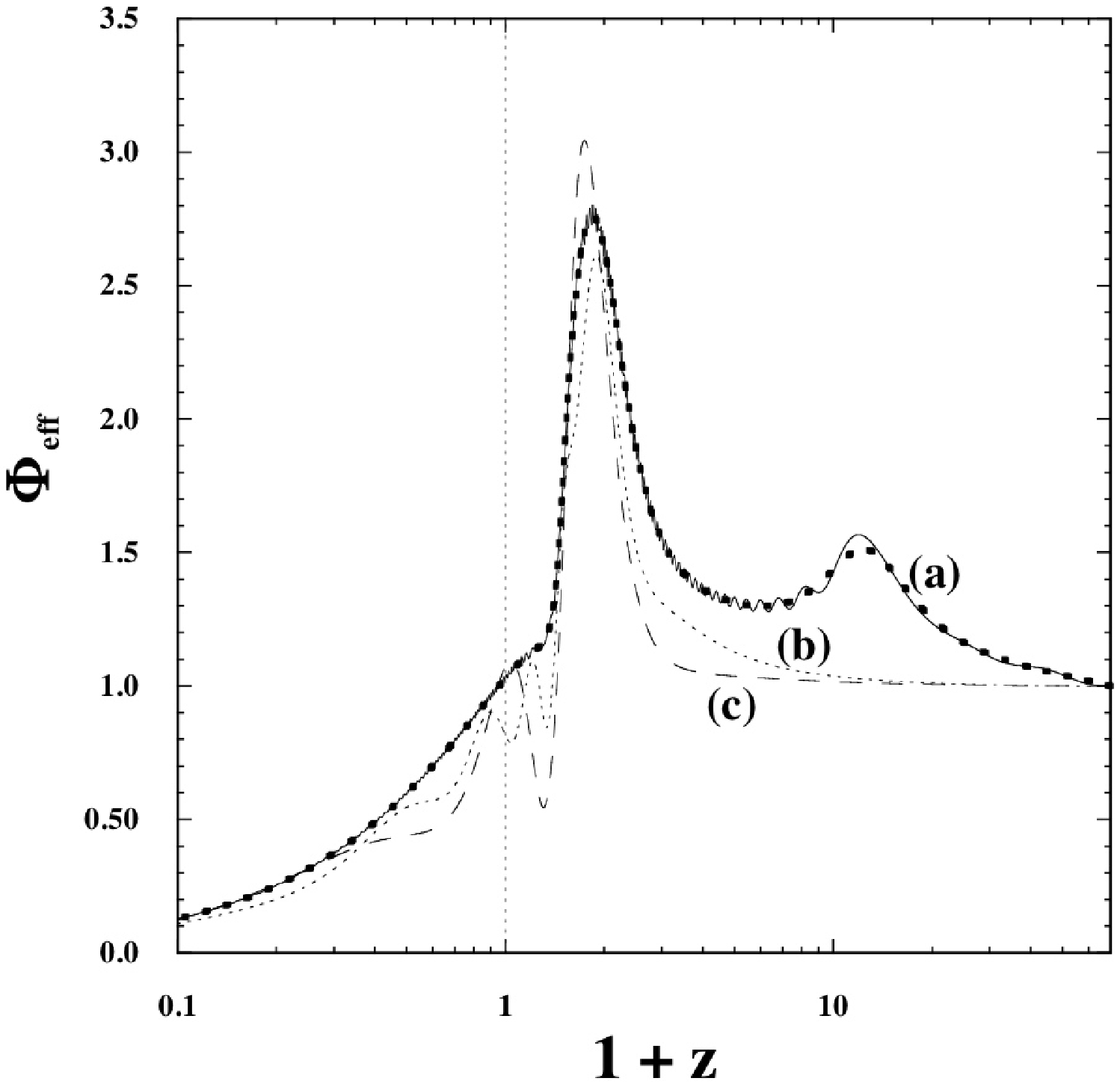} 
\par\end{centering}
\caption{Evolution of the perturbations
for $\alpha=1.37$ and $\beta=0.44$ with the 
background initial conditions $r_1=0.03$ and $r_2=0.003$
(corresponding to the case (A) of Fig.~\ref{wde}).
(Left) $\delta_m/a$ versus $z$ for the wave numbers
(a) $k=300a_0H_0$, (b) $k=30a_0H_0$, and (c) $k=5a_0H_0$.
(Right) $\Phi_{\rm eff}$ versus $z$ for the wave numbers
(a) $k=300a_0H_0$, (b) $k=10a_0H_0$, and (c) $k=5a_0H_0$.
Note that $\delta_m/a$ and $\Phi_{\rm eff}$ are divided by 
their initial amplitudes $\delta_m (t_i)/a(t_i)$ and 
$\Phi_{\rm eff}(t_i)$, respectively, so that their initial 
values are normalized to be 1.
The bold dotted lines show the results obtained under the 
quasistatic approximation on subhorizon scales.
The choice of initial conditions for perturbations 
is explained in the text.}
\centering{}\label{delmPhifig} 
\end{figure}

\begin{figure}
\begin{centering}
\includegraphics[width=3.3in,height=3.3in]{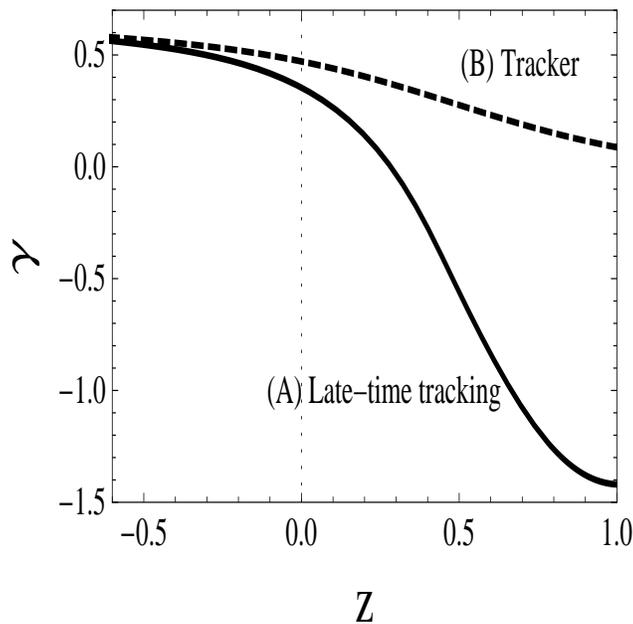} 
\par\end{centering}
\caption{Variation of the growth index $\gamma$ of matter 
perturbations for $\alpha=1.37$ and $\beta=0.44$ with the mode $k=300a_0H_0$. 
The initial conditions of $r_1$ and $r_2$ for the cases 
(A) and (B) are the same as those given in the caption 
of Fig.~\ref{wde}.}
\centering{}\label{gamma} 
\end{figure}

%
\subsection{Case of late-time tracking solutions}

Let us consider the evolution of perturbations for the 
late-time tracking solutions.
In Fig.~\ref{delmPhifig} we plot $\delta_m/a$ and 
$\Phi_{\rm eff}$ versus the redshift $z$ with the 
model parameters and initial conditions corresponding 
to the case (A) of Fig.~\ref{wde}.
For the mode $k=300a_0H_0$ we find that the full 
numerical result shows excellent agreement with that obtained
under the quasistatic approximation on subhorizon scales.
The difference starts to appear for the modes 
$k/(a_0H_0)<{\cal O}(10)$.
The left panel of Fig.~\ref{delmPhifig} shows that, 
on larger scales, the growth of $\delta_m$
tends to be less significant.
For the modes $k \gg a_0H_0$ the matter perturbation 
evolves faster than $a$ during the matter era 
(i.e. faster than in the case of GR, $\delta_m \propto a$).
The growth of $\delta_m/a$ turns into decrease 
after the Universe enters the epoch of cosmic acceleration.

In Fig.~\ref{gamma} we show the evolution of the growth index 
$\gamma$ defined by \cite{Peebles}
\be
\delta_m'/\delta_m=(\Omega_m)^{\gamma}\,.
\label{gammadef}
\ee
The case (A) in Fig.~\ref{gamma} corresponds to the same model parameters and 
initial conditions as those given in the numerical simulations 
of Fig.~\ref{delmPhifig}.
Unlike the $\Lambda$CDM model in which $\gamma$ is nearly constant 
($\simeq 0.55$ \cite{Wang}) in the low-redshift regime $0<z<1$, 
the variation of $\gamma$ is significant for the solutions 
that approach the tracker at late times.
Moreover the growth index today for the mode $k=300a_0H_0$
is $\gamma_0=0.35$, which is quite different
from that in the $\Lambda$CDM. 
For the wave lengths relevant to large-scale structures we find that 
the growth index today exhibits almost no dispersion with respect
to $k$. This property is different from viable $f(R)$ dark energy 
models in which $\gamma_0$ can be dispersed \cite{TGMP}.
If we choose smaller initial values of $r_1$ (i.e.
the later tracking), we find that $\gamma_0$ tends to 
be smaller, e.g. $\gamma_0=0.23$ for the initial 
conditions $r_1=0.01$, $r_2=0.003$ with the model 
parameters $\alpha=1.37$, 
$\beta=0.44$.\footnote{For the dark energy models with constant 
$w_{\rm DE}$ and $\gamma$, the current observations still allow
the large parameter space of $\gamma$ ranging from 
0.2 to 0.6. \cite{Rapetti}.}

The right panel of Fig.~\ref{delmPhifig} shows that,  
unlike the $\Lambda$CDM model, the effective 
gravitational potential $\Phi_{\rm eff}$ 
changes in time even during the matter era
for the modes $k \gg aH$. 
This can be understood as follows.
For the initial conditions corresponding to the late-time
tracking solutions the effective gravitational coupling 
$G_{\rm eff}$ and the anisotropic parameter $\eta$ are given by 
Eqs.~(\ref{Geffap}) and (\ref{etaap}), respectively,
before reaching the tracker.
Since $G_{\rm eff}/G \simeq 1+255\beta r_2/8>1$ and 
$\eta \simeq 1+129\beta r_2/8>1$ in this regime, 
we have the larger growth rates of $\Phi_{\rm eff}$ 
and $\delta_m$ relative to those in GR.
In particular the term $(G_{\rm eff}/G)(1+\eta)/2$
in Eq.~(\ref{Phieffap}) is larger than 1, which leads to 
the additional enhancement of $\Phi_{\rm eff}$ to the growth 
coming from $\delta_m$.
Note that in $f(R)$ gravity \cite{quasi2} and in 
Brans-Dicke theory \cite{Bransper}
one has $(G_{\rm eff}/G)(1+\eta)/2=1$, so that the 
evolution of $\delta_m$ is directly related with  
the variation of $\Phi_{\rm eff}$.
In Galileon gravity the unusual behavior of the 
anisotropic parameter $\eta$ leads to the nontrivial 
evolution of perturbations.
For the model parameters $\alpha=1.37$ and $\beta=0.44$, 
Eq.~(\ref{Geffde}) gives $G_{\rm eff} \simeq 0.68 G$
at the dS fixed point.
As we see in Fig.~\ref{delmPhifig}, $\Phi_{\rm eff}$ 
begins to decrease at some point after the matter era.

For the large-scale modes relevant to the ISW effect in 
CMB anisotropies, i.e. $k/(a_0H_0) \lesssim 10$,
the effective gravitational potential is nearly 
constant in the early matter-dominated epoch, 
see the cases (b) and (c) 
in the right panel of Fig.~\ref{delmPhifig}.
However, $\Phi_{\rm eff}$ exhibits temporal growth 
during the transition from the matter era 
to the epoch of cosmic acceleration.
Note that in the $\Lambda$CDM model $\Phi_{\rm eff}$ 
decays without the temporal growth after the matter era.
The characteristic variation of $\Phi_{\rm eff}$ in 
the Galileon model should leave observational signatures
in CMB anisotropies as the ISW effect. 

The above numerical results correspond to the fixed 
values of $\alpha$ and $\beta$ ($\alpha=1.37$, $\beta=0.44$).
If we use the bounds $\alpha=1.404 \pm 0.057$, 
$\beta=0.419 \pm 0.023$ constrained by Union2$+$CMB$+$BAO
data sets, the effective gravitational coupling at the dS point 
is restricted in the range $0.5G<G_{\rm eff}<0.72G$. 
The bounds coming from Constitution$+$CMB$+$BAO data give 
the similar constraint, i.e. $0.5G<G_{\rm eff}<0.71G$.
For the model parameters close to the upper limit 
$\alpha=2\beta+2/3$ of the allowed parameter space
(i.e. $G_{\rm eff}$ is close to $0.5G$ at the 
dS point), the parameter $\eta$ tends to show 
a divergence during the transition from the matter 
era to the dS epoch. 
If $G_{\rm eff}$ is larger than $0.66G$, we find that 
such divergent behavior can be typically avoided.

For the parameters $\alpha$ and $\beta$ constrained 
observationally, the values of $\gamma_0$ are usually 
less than 0.4 for the late-time tracking with the 
minimum values of $w_{\rm DE}$ larger than $-1.3$.
In addition the later tracking leads to smaller
values of $\gamma_0$ with larger variations of $\gamma$.
This property of the Galileon model can be clearly 
distinguished from the $\Lambda$CDM model. 

\subsection{Case of early-time tracking solutions}

\begin{figure}
\begin{centering}
\includegraphics[width=3.3in,height=3.2in]{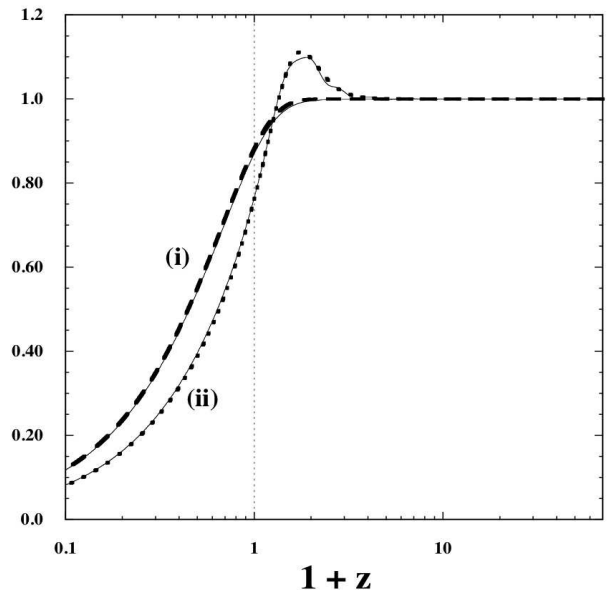} 
\par\end{centering}
\caption{Evolution of (i) $\delta_m/a$ and (ii) $\Phi_{\rm eff}$
versus $z$ for $\alpha=1.37$ and $\beta=0.44$ with the 
initial conditions $r_1=0.999$ and $r_2=7.0 \times 10^{-11}$
[corresponding to the case (B) of Fig.~\ref{wde}].
Both $\delta_m/a$ and $\Phi_{\rm eff}$ are divided by 
their initial amplitudes.
In this case the background cosmological solution is 
on the tracker from the onset of integration.
The solid line shows the evolution of perturbations 
for the mode $k=5a_0H_0$, whereas the bold dotted line and bold dashed line 
represent the result derived under the quasistatic 
approximation on subhorizon scales.}
\centering{}\label{Phieff2} 
\end{figure}

If the solutions are already close to the tracker 
in the early matter-dominated epoch, the evolution 
of perturbations is different from that discussed above. 
For the tracker one has $r_2 \propto a^6$ 
during the matter dominance, 
which is much faster than the evolution $r_2 \propto a^{3/8}$
in the regime $r_1 \ll 1$ and $r_2 \ll 1$.
This means that $r_2$ is very much smaller than 1
for the redshift $z \gtrsim 1$.
Then the second terms on the r.h.s. of Eqs.~(\ref{Gefftracker}) 
and (\ref{etatracker}) are suppressed relative to the 
first terms until recently. 
In Fig.~\ref{Phieff2} we plot the variation of $\delta_m/a$
and $\Phi_{\rm eff}$ for $\alpha=1.37$ and $\beta=0.44$
with the background initial conditions corresponding to
the case (B) in Fig.~\ref{wde}.
The solid line shows the evolution of perturbations
for the mode $k=5a_0H_0$, whereas the bold dotted line 
and bold dashed line represent the result 
derived under the quasistatic approximation on subhorizon scales.
In this case the evolution of the large-scale mode with $k=5a_0H_0$ 
is similar to that for the modes deep inside the Hubble radius.

Figure \ref{Phieff2} shows that the growth of matter 
perturbations in the deep matter era is almost identical 
to that in the $\Lambda$CDM model.
In the regime $r_1 \simeq 1$ and $r_2 \ll 1$ one has 
$G_{\rm eff}/G \simeq 1+3.2r_2$ and
$\eta \simeq 1-3.7r_2$, both of which are very close to 1.
Recall that at the dS point $G_{\rm eff}/G \simeq 0.68$ and $\eta=1$.
Since $\eta$ is very close to 1 and $G_{\rm eff}$
is different from $G$ only around the dS solution,  
the evolution of perturbations is milder than that 
corresponding to the late-time tracking solutions.
The case (B) of Fig.~\ref{gamma} shows $\gamma$
versus $z$ for the mode $k=300a_0H_0$ with the initial 
condition corresponding to the tracker solution.
In this case the growth index today is found to be 
$\gamma_0 \simeq 0.47$ with the variation of 
$\gamma$ in the low-redshift regime.
In Fig.~\ref{Phieff2} we also find the small temporal growth 
of $\Phi_{\rm eff}$ from the end of the matter era to 
the epoch of cosmic acceleration. 
This can also give rise to a nonnegligible contribution 
to the ISW effect in CMB anisotropies.
We recall, however, that the tracker solution is not favored
at the background level by the joint analysis of
observational data.

\subsection{Case of $\alpha=\beta=0$}

Let us finally consider the model in which only the terms 
up to ${\cal L}_3$ are present, i.e. $\alpha=\beta=0$.
Although such a model is disfavored observationally, 
it is of interest to study what happens for the evolution of 
perturbations due to the peculiar divergence of $G_{\rm eff}$ 
at the dS point (which occurs along 
the line $\alpha=2\beta$).

First of all we have numerically found that the late-time 
tracking behavior as in the case (A) of Fig.~\ref{wde}
is not easy to be realized even by choosing many different 
initial conditions. 
This is the main reason why the combined data analyses 
at the background level do not favor such model parameters.
For $\alpha=\beta=0$ one has $\eta=1$
independent of the values of $r_1$ and $r_2$,
whereas $G_{\rm eff} \simeq G$ for $r_1 \ll 1, r_2 \ll 1$
and $G_{\rm eff} \simeq 1+r_2/4$ for $r_1=1, r_2 \ll 1$.

\begin{figure}
\begin{centering}
\includegraphics[width=3.2in,height=3.1in]{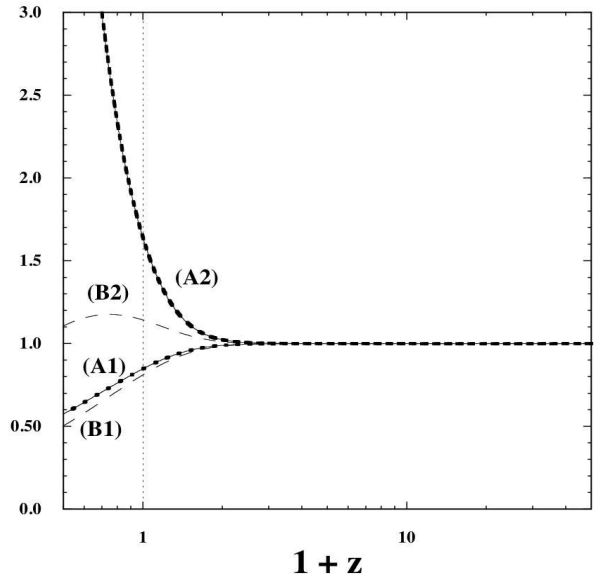} 
\par\end{centering}
\caption{Evolution of perturbations for $\alpha=0$ and $\beta=0$ 
with the initial conditions $r_1=0.05$ and $r_2=0.001$.
The lines correspond as follows: (A1) $\delta_m/a$ for the mode
$k=300a_0H_0$, (A2) $\Phi_{\rm eff}$ for the mode
$k=300a_0H_0$, (B1) $\delta_m/a$ for the mode
$k=5a_0H_0$, and (B2) $\Phi_{\rm eff}$ for the mode
$k=5a_0H_0$. Both $\delta_m/a$ and $\Phi_{\rm eff}$ are 
divided by their initial amplitudes.
The bold dotted line and bold dashed line correspond to 
the results obtained under the quasistatic approximation 
on subhorizon scales.}
\centering{}\label{albezero} 
\end{figure}

In Fig.~\ref{albezero} we plot the variation of $\delta_m/a$
and $\Phi_{\rm eff}$ with the initial conditions 
$r_1=0.05$ and $r_2=0.001$ for two different wave numbers:
$k=300a_0H_0$ and $k=5a_0H_0$. In this simulation 
the solution reaches the tracker region ($r_1 \simeq 1$) 
around the redshift $z=5$, e.g. $r_1=0.99$ and 
$r_2=4.14 \times 10^{-4}$ at $z=5.34$.
Hence the evolution of perturbations for $z \gtrsim 5$ is 
similar to that in the $\Lambda$CDM model.
The effective gravitational coupling $G_{\rm eff}$ increases
with the growth of $r_2$ for $z \lesssim 5$.
In this case the numerical value of $G_{\rm eff}$ today is 
$G_{\rm eff} \simeq 1.94 G$ with $r_2 \simeq 0.72$.
For the mode $k=300a_0H_0$ the growth index today is
found to be $\gamma_0 \simeq 0.43$, which is smaller 
than the value in the $\Lambda$CDM model.

In Fig.~\ref{albezero} the effective gravitational potential 
for the mode $k=300a_0H_0$ grows around the present epoch 
in spite of the decrease of $\delta_m/a$.
This can be understood by the fact that the term 
$(G_{\rm eff}/G)(1+\eta)/2$, which is equivalent to 
$G_{\rm eff}/G$ for $\alpha=\beta=0$, continues to 
grow toward the dS solution.
While the decrease of the density parameter $\Omega_m$
overwhelms the increase of $G_{\rm eff}/G$ in Eq.~(\ref{delmeqd}) 
for matter perturbations, the presence of the additional term $(aH/k)^2$
in Eq.~(\ref{Phieffap}) stimulates the growth of $\Phi_{\rm eff}$  
around today. Numerically we find that the growth of 
$\Phi_{\rm eff}$ eventually turns into the decrease 
in future as $\delta_m$ decays
sufficiently (which is not shown in Fig.~\ref{albezero}).
For $k=5a_0H_0$ the growth of $\Phi_{\rm eff}$
around today is milder relative to the modes
deep inside the Hubble radius.

The anticorrelation between $\delta_m/a$ and $\Phi_{\rm eff}$ 
seen in Fig.~\ref{albezero} is similar to the one found 
in the generalized Galileon model \cite{Kobayashi1}.
This anticorrelation mainly comes from the fact that 
$G_{\rm eff}/G$ continues to grow toward the dS point 
and that the parameter $\eta$ does not become smaller than 1
to compensate the growth of $G_{\rm eff}/G$ 
(unlike $f(R)$ gravity and Brans-Dicke theory).
In the numerical simulations of Fig.~\ref{delmPhifig} the 
effective gravitational coupling is larger than $G$ 
in the regime $r_1 \ll 1$ and $r_2 \ll 1$, but 
it finally approaches the asymptotic value 
$G_{\rm eff} \simeq 0.68 G$.
In this case both $\delta_m/a$ and $\Phi_{\rm eff}$
start to decay by today, 
whose property is different from that in the case  
$\alpha=\beta=0$.
Since the observational constraints on $\alpha$ and $\beta$ 
do not allow the region close to the line $\alpha=2\beta$, 
the anticorrelation between $\delta_m/a$ and $\Phi_{\rm eff}$
does not typically occur for the viable model parameters.

\section{Conclusions}
\label{secconclude}

We have studied the dynamics of cosmological perturbations in 
the Galileon model whose Lagrangian satisfies the Galileon symmetry 
$\partial_{\mu}\phi \to \partial_{\mu}\phi+b_{\mu}$
in the flat space-time.
In this theory there exists stable dS solutions with 
$\dot{\phi}=$~constant.
In the deep matter era the General Relativistic
behavior can be recovered because of a small field velocity, 
but the Universe finally enters the epoch of cosmic acceleration
after the growth of $\dot{\phi}$.

Before reaching the dS attractor, the solutions approach 
a tracker characterized by $r_1=1$
(i.e. the evolution $\dot{\phi} \propto H^{-1}$).
The tracking epoch depends on the initial conditions of the background
cosmological variables. The combined data analysis based on the 
background expansion history of the Universe 
(SN Ia$+$CMB shift parameters$+$BAO) favor the late-time 
tracking around today. This corresponds to the initial conditions 
$r_1 \ll 1$ and $r_2 \ll 1$ in the deep matter era, so that 
$r_1$ approaches 1 in the low-redshift regime ($z \lesssim 1$).

The study about the evolution of cosmological perturbations can allow 
us to distinguish the Galileon model from the $\Lambda$CDM model
further. In the presence of a nonrelativistic perfect fluid,
we have derived full perturbation equations for the model 
described by the action (\ref{action}).
In spite of their complexities it is possible to obtain simpler equations
for the matter perturbation $\delta_m$ and the effective 
gravitational potential $\Phi_{\rm eff}$ under a quasistatic 
approximation on subhorizon scales.

The two important quantities associated with the growth of $\delta_m$
and $\Phi_{\rm eff}$ are the effective gravitational coupling $G_{\rm eff}$
and the anisotropic parameter $\eta=-\Phi/\Psi$.
In three distinct regimes we have approximately derived the expressions 
of $G_{\rm eff}$ and $\eta$, see Eqs.~(\ref{Geffap})-(\ref{etade}).
For the initial conditions that lead to the late-time tracking behavior,
the solutions start from the regime $r_1 \ll 1$, $r_2 \ll 1$
with positive $\beta$ (which is required to avoid ghosts).
In this regime Eqs.~(\ref{Geffap}) and (\ref{etaap}) show that 
$G_{\rm eff}/G \simeq 1+255 \beta r_2/8>1$ and 
$\eta \simeq 1+129 \beta r_2/8>1$.
This gives rise to the larger growth rates of $\delta_m$ and 
$\Phi_{\rm eff}$ relative to those in the $\Lambda$CDM model.

In $f(R)$ gravity and Brans-Dicke theory the anisotropic parameter 
$\eta$ is less than 1, so that the combination 
$(G_{\rm eff}/G)(1+\eta)/2$ remains to be 1.
In the Galileon model the fact that $\eta>1$ in the regime 
$r_1 \ll 1$, $r_2 \ll 1$ leads to the further growth of $\Phi_{\rm eff}$
for subhorizon modes. At the dS fixed point we have that 
$G_{\rm eff}/G=1/[3(\alpha-2\beta)]$ and $\eta=1$.
For the parameters of $\alpha$ and $\beta$ observationally constrained
at the background level \cite{NFT},the effective gravitational 
coupling $G_{\rm eff}$ is smaller than $0.72G$. 
Then the perturbations $\delta_m/a$ and $\Phi_{\rm eff}$
turn into decrease around today, as we see in the numerical 
simulation of Fig.~\ref{delmPhifig}.

For the solutions that approach the tracker at the early epoch 
of the matter era, $G_{\rm eff}$ and $\eta$ are approximately 
given by Eqs.~(\ref{Gefftracker}) and (\ref{etatracker}), respectively, 
in the regime $r_1=1, r_2 \ll 1$.
Since the evolution of $r_2$ in the tracking regime is very fast 
($r_2 \propto a^6$), the correction term $\beta r_2$ to 
$G_{\rm eff}$ and $\eta$ becomes important around 
the present epoch in which $r_2$ grows to the order of 1.
Hence the deviation from the $\Lambda$CDM model for 
the evolution of $\delta_m$ and $\Phi_{\rm eff}$
appears only at late times, as we see in Fig.~\ref{Phieff2}.
The early tracking behavior is however disfavored from 
joint observational constraints at the background level.

For the late-time tracking solutions we have found that
the growth index $\gamma$ of matter perturbations
rapidly changes in the low-redshift 
regime [as in the case (A) of Fig.~\ref{gamma}].
The values of $\gamma$ today on the scales relevant to 
large-scale structures are typically smaller than 0.4
for the model parameters constrained by the 
observations at the background level.
As the tracking occurs at a later epoch, the presents values 
of $\gamma$ tend to be smaller.
On large scales relevant to the ISW effect in CMB anisotropies
the effective gravitational potential $\Phi_{\rm eff}$
exhibits a temporal growth around the present epoch.
These properties can potentially restrict the allowed region 
constrained by the background cosmic
expansion history further.
It will be of interest to constrain the Galileon model 
from the combined data analysis of large-scale structures, 
CMB, and weak lensing.

\section*{ACKNOWLEDGEMENTS}
The work of A.\,D.\,F.\ and S.\,T.\ was supported by the
Grant-in-Aid for Scientific Research Fund of the 
JSPS Nos.~10271 and 30318802.
S.\,T.\ also thanks financial support for the Grant-in-Aid for
Scientific Research on Innovative Areas (No.~21111006). 
S.\,T.\ is grateful to Bin Wang for warm hospitality 
during his stay in Shanghai Jiao Tong University.

\appendix

\section{Bianchi identities}\label{sec:bianchi}

We confirm the consistency of perturbation equations by using 
the Bianchi identities. We write the action (\ref{action}) in the form 
\be
S=\int {\rm d}^{4}x\sqrt{-g}\,{\cal L}\,,
\label{action2}
\ee
where ${\cal L}$ is the total Lagrangian. 
Variation of the action (\ref{action2}) with respect to 
$g_{\alpha \beta}$ gives
\be
\delta S=\int {\rm d}^{4}x\sqrt{-g}\,
\Sigma^{\alpha\beta}\delta g_{\alpha\beta}\,.
\ee
Then the Bianchi identities lead to 
\be
\label{Bianchi}
\nabla_{\beta}\Sigma^{\alpha\beta}=\partial_{\beta}\Sigma^{\alpha\beta}
+\Gamma_{\lambda\beta}^{\alpha}\Sigma^{\lambda\beta}
+\Gamma_{\lambda\beta}^{\beta}\Sigma^{\alpha\lambda}=0\,,
\ee
where $\Gamma_{\lambda\beta}^{\alpha}$ is the Christoffel symbol.

We perturb Eq.~(\ref{Bianchi}) at first order. Then we find
\be
\partial_{\beta}\delta\Sigma^{\alpha\beta}+
\Gamma_{\lambda\beta}^{\alpha}\delta\Sigma^{\lambda\beta}
+\Gamma_{\lambda\beta}^{\beta}\delta\Sigma^{\alpha\lambda}=0\,,
\ee
where we have used the fact that the background equation
satisfies $\Sigma^{\alpha \beta}=0$.
For the case $\alpha=0$ we have 
\be
\partial_{0}\delta\Sigma^{00}+\partial_{i}\delta\Sigma^{0i}
+\Gamma_{\lambda\beta}^{0}\delta\Sigma^{\lambda\beta}
+\Gamma_{0\beta}^{\beta}\delta\Sigma^{00}
+\Gamma_{i\beta}^{\beta}\delta\Sigma^{0i}=0\,.
\label{Gamma}
\ee
Using the relations $\Gamma_{00}^{0}=0$, $\Gamma_{0i}^{0}=0$,
$\Gamma_{ij}^{0}=a^{2}H\delta_{ij}$,
$\Gamma_{00}^{i}=0$, $\Gamma_{0j}^{i}=H\delta_{ij}$,
$\Gamma_{jk}^{i}=0$, Eq.~(\ref{Gamma}) reduces to 
\be
\partial_{0}\delta\Sigma^{00}+\partial_{i}\delta\Sigma^{0i}
+a^{2}H\sum_i\delta\Sigma^{ii}+3H\delta\Sigma^{00}=0\,.
\label{delbi}
\ee

In our Galileon model the condition (\ref{delbi}) 
leads to the following relation
\be
\label{eq:bianchi1}
E_1\equiv\dot E_\Psi+3H E_\Psi-H E_\Phi
-\frac{k^2}{a^2}\,E_\chi=0\,,
\ee
where $E_\Psi$ corresponds to Eq.~(\ref{eq:Psi}), $E_\Phi$ 
to Eq.~(\ref{eq:Phi}), 
and $E_\chi$ to Eq.~(\ref{eq:chi}).
Combining  Eq.~(\ref{eq:delphi}) with Eq.~(\ref{eq:bianchi1})
and inserting the background equations of motion, we find
\begin{equation}
\label{eq:bianchi2}
E_1-\dot{\phi} E_{\delta\phi}=
\rho_m \left(\frac{k^2}{a^2}\,\chi-3\dot\Phi-\frac{k^2}{a^2}\,v
-\dot\delta\right)=0\,,
\end{equation}
which matches with Eq.~(\ref{delrhoeq1}) for $w=0$.

Equation (\ref{eq:constr2}) can be also derived from the spatial part of 
the Bianchi identities. It follows that 
\be
\label{eq:constr2B}
\dot E_\chi+3H E_\chi-\frac13 E_\Phi
=\frac13 \frac{k^2}{a^2}\tilde{E}_\gamma=0\,,
\ee
where $\tilde{E}_\gamma$ corresponds to Eq.~(\ref{eq:constr2}).

\section{Coefficients of perturbation equations}
\label{coefficients}

In this Appendix we show the coefficients of the perturbation 
equations (\ref{eq:Psi}), (\ref{eq:Phi}), (\ref{eq:chi}),  
and (\ref{eq:delphi}).

\flushleft{\underline{Eq.~(\ref{eq:Psi})}}
\ba
& &A_1=6HM_{\rm pl}^2-3c_3 \dot{\phi}^{3}/M^3
+45c_4H \dot{\phi}^{4}/M^6
-63c_5{H}^{2}\dot{\phi}^{5}/M^{9}\,,\\
& &A_2=c_2 \dot{\phi}-9c_3H \dot{\phi}^2/M^3
+90c_4 H^2 \dot{\phi}^3/M^6
-105c_5 H^3 \dot{\phi}^4/M^9\,,\\
& &A_3=2M_{\rm pl}^2+3c_4 \dot{\phi}^4/M^6-6 c_5 H \dot{\phi}^5/M^9\,,\\
& &A_4=2\rho_m-9H^2M_{\rm pl}^2
-c_1M^3 \phi/2-3c_2 \dot{\phi}^2/2+15c_3 H \dot{\phi}^3/M^3
-315c_4 H^2 \dot{\phi}^4/(2M^6)+189 c_5 H^3 \dot{\phi}^5/M^9\,,\\
& &A_5=-2HM_{\rm pl}^2+c_3 \dot{\phi}^3/M^3-15c_4 H \dot{\phi}^4/M^6
+21c_5 H^2 \dot{\phi}^5/M^9\,,\\
& &A_6=-c_3 \dot{\phi}^2/M^3+12 c_4 H\dot{\phi}^3/M^6
-15c_5 H^2 \dot{\phi}^4/M^9\,.
\ea
\vspace{0.3cm}
\underline{Eq.~(\ref{eq:Phi})}
\ba
& &B_1=3A_3\,,\\
& &B_2=3A_6\,,\\
& &B_3=18H M_{\rm pl}^2+27c_4 H \dot{\phi}^4/M^6
+36c_4 \dot{\phi}^3 \ddot{\phi}/M^6-54c_5 H^2 \dot{\phi}^5/M^9
-90c_5 H \dot{\phi}^4 \ddot{\phi}/M^9-18c_5 \dot{H} \dot{\phi}^5/M^9
\,,\\
& &B_4=-3c_2 \dot{\phi}-6c_3 \dot{\phi} \ddot{\phi}/M^3+
36c_4 \dot{H} \dot{\phi}^3/M^6+108c_4 H \dot{\phi}^2 \ddot{\phi}/M^6
+54 c_4 H^2 \dot{\phi}^3/M^6 \nonumber \\
& &~~~~~~~\,-90c_5 H^3 \dot{\phi}^4/M^9
-180c_5 H^2 \dot{\phi}^3 \ddot{\phi}/M^9
-90c_5 H \dot{H} \dot{\phi}^4/M^9 \,,\\
& &B_5=-A_1\,,\\
& &B_6=2M_{\rm pl}^2-c_4 \dot{\phi}^4/M^6
-6c_5 \dot{\phi}^4 \ddot{\phi}/M^9\,,\\
& &B_7=12c_4 \dot{\phi}^2 \ddot{\phi}/M^6+4c_4 H\dot{\phi}^3/M^6
-6c_5 \dot{H}\dot{\phi}^4/M^9-6c_5 H^2 \dot{\phi}^4/M^9
-24 c_5 H \dot{\phi}^3 \ddot{\phi}/M^9\,,\\
& &B_8=A_3\,,\\
& &B_9=-9H^2M_{\rm pl}^2-6\dot{H}M_{\rm pl}^2-3\rho_m
+3c_1 M^3 \phi/2+3c_2 \dot{\phi}^2/2+9c_3 \dot{\phi}^2 \ddot{\phi}/M^3 
-135c_4H^2 \dot{\phi}^4/(2M^6)
\nonumber \\
& &~~~~~~~\,-180c_4 H\dot{\phi}^3 \ddot{\phi}/M^6
-45c_4 \dot{H} \dot{\phi}^4/M^6
+315c_5 H^2 \dot{\phi}^4 \ddot{\phi}/M^9
+126c_5 H^3 \dot{\phi}^5/M^9
+126c_5 H \dot{H} \dot{\phi}^5/M^9\,,\\
& &B_{10}=-A_3\,,\\
& &B_{11}=-2H M_{\rm pl}^2-3c_4H \dot{\phi}^4/M^6
-12c_4 \dot{\phi}^3 \ddot{\phi}/M^6
+6c_5 \dot{H} \dot{\phi}^5/M^9+30c_5 H \dot{\phi}^4 
\ddot{\phi}/M^9+6c_5 H^2 \dot{\phi}^5/M^9\,.
\ea
\vspace{0.3cm}
\underline{Eq.~(\ref{eq:chi})}
\ba
& &C_1=A_3\,,\qquad C_2=A_6\,,\qquad C_3=-A_1/3\,,\\
& &C_4=-c_2 \dot{\phi}+3c_3 H \dot{\phi}^2/M^3
-18c_4 H^2 \dot{\phi}^3/M^6+15c_5 H^3 \dot{\phi}^4/M^9\,.
\ea
\vspace{0.3cm}
\underline{Eq.~(\ref{eq:delphi})}
\ba
& &D_1=3A_6\,,\\
& &D_2=c_2-6c_3 H\dot{\phi}/M^3+54c_4 H^2 \dot{\phi}^2/M^6
-60c_5 H^3 \dot{\phi}^3/M^9\,,\\
& &D_3=3c_2 \dot{\phi}-18c_3 H \dot{\phi}^2/M^3-6c_3 
\dot{\phi} \ddot{\phi}/M^3+108c_4 H \dot{\phi}^2 \ddot{\phi}/M^6
+162 c_4 H^2 \dot{\phi}^3/M^6+36c_4 \dot{H} \dot{\phi}^3/M^6 
\nonumber \\ 
& &~~~~~~~~-90c_5 (H \dot{H} \dot{\phi}^4/M^9
+2H^3 \dot{\phi}^4/M^9+2H^2 \dot{\phi}^3
\ddot{\phi}/M^9)\,,\\
& &D_4=3c_2H-6c_3 \dot{H} \dot{\phi}/M^3-18c_3 H^2 \dot{\phi}/M^3
-6c_3 H\ddot{\phi}/M^3+162c_4 H^3 \dot{\phi}^2/M^6+108c_4 H \dot{H}
\dot{\phi}^2/M^6 \nonumber \\
& &~~~~~~~~+108c_4 H^2 \dot{\phi} \ddot{\phi}/M^6-180c_5
(H^3 \dot{\phi}^2 \ddot{\phi}/M^9+H^2 \dot{H} \dot{\phi}^3/M^9
+H^4 \dot{\phi}^3/M^9)\,,\\ 
& &D_5=-A_2\,,\\
& &D_6=-A_6\,,\\
& &D_7=B_7\,,\\
& &D_8=3c_1M^3/2+3c_2 \ddot{\phi}+9c_2 H \dot{\phi}
-27c_3 H^2 \dot{\phi}^2/M^3-18c_3 H\dot{\phi} \ddot{\phi}/M^3
-9c_3 \dot{H} \dot{\phi}^2/M^3+162 c_4 H^2 \dot{\phi}^2 \ddot{\phi}/M^6 
\nonumber \\
& &~~~~~~~~+162c_4 H^3 \dot{\phi}^3/M^6+108c_4 H \dot{H} 
\dot{\phi}^3/M^6-180c_5H^3 \dot{\phi}^3 \ddot{\phi}/M^9
-135c_5(H^2 \dot{H} \dot{\phi}^4/M^9+H^4 \dot{\phi}^4/M^9) \,,\\
& &D_9=c_2-4c_3 H \dot{\phi}/M^3
-2c_3 \ddot{\phi}/M^3+26c_4 H^2 \dot{\phi}^2/M^6
+12c_4 \dot{H} \dot{\phi}^2/M^6
+24c_4 H \dot{\phi} \ddot{\phi}/M^6 \nonumber \\
& &~~~~~~~~-36c_5 H^2 \dot{\phi}^2 \ddot{\phi}/M^9
-24c_5 (H^3 \dot{\phi}^3/M^9
+H \dot{H} \dot{\phi}^3/M^9)\,,\\
& &D_{10}=A_6\,,\\
& &D_{11}=c_1M^3/2-3c_2 H\dot{\phi}-c_2 \ddot{\phi}
+27c_3 H^2 \dot{\phi}^2/M^3+18c_3 H \dot{\phi} \ddot{\phi}/M^3
+9c_3 \dot{H} \dot{\phi}^2/M^3-180c_4 H \dot{H} \dot{\phi}^3/M^6
\nonumber \\
& &~~~~~~~~-270c_4 (H^2 \dot{\phi}^2 \ddot{\phi}/M^6+H^3 \dot{\phi}^3/M^6)
+420 c_5 H^3 \dot{\phi}^3 \ddot{\phi}/M^9
+315c_5 (H^4 \dot{\phi}^4/M^9+H^2 \dot{H} \dot{\phi}^4/M^9)\,,\\
& &D_{12}=-c_2 \dot{\phi}+4c_3 H\dot{\phi}^2/M^3+2c_3 \dot{\phi}
\ddot{\phi}/M^3-30c_4 H^2 \dot{\phi}^3/M^6-36c_4 H\dot{\phi}^2 
\ddot{\phi}/M^6-12c_4 \dot{H} \dot{\phi}^3/M^6 \nonumber \\
& &~~~~~~~~\,+30c_5 (H \dot{H} \dot{\phi}^4/M^9+H^3 \dot{\phi}^4/M^9
+2H^2 \dot{\phi}^3 \ddot{\phi}/M^9)\,.
\ea
%


\end{document}